\documentclass[twocolumn,longbibliography,floatfix,tightline]{revtex4-1}
\usepackage{amssymb}
\usepackage{amsmath}
\usepackage{colordvi}
\usepackage[colorlinks]{hyperref}
\usepackage{amsthm}
\usepackage{subeqnarray}
\usepackage{cases}
\usepackage{enumerate}
\usepackage{graphicx}
\usepackage{epsfig}
\usepackage{epstopdf}
\usepackage{subfigure}
\usepackage{color}
\usepackage{bm}

\def\avg(#1){\langle#1\rangle}

\def\be{\begin{equation}}
\def\ee{\end{equation}}
\def\bea{\begin{eqnarray}}
\def\eea{\end{eqnarray}}

\def\nn{\nonumber}

\begin{document}

\title{Synthetic Landau levels and spinor vortex matter on Haldane spherical surface with magnetic monopole}
\author{Xiang-Fa Zhou$^{1,2}$, Congjun Wu$^3$, Guang-Can Guo$^{1,2}$, Ruquan Wang$^{4,5}$}
\author{Han Pu$^{6,7}$}
\email{hpu@rice.edu}
\author{Zheng-Wei Zhou$^{1,2}$}
\email{zwzhou@ustc.edu.cn}
\affiliation{$^1$Key Laboratory of Quantum Information, Chinese Academy of Sciences, University of Science and Technology of China, Hefei, 230026, China \\
$^2$Synergetic Innovation Center of Quantum Information
and Quantum Physics, University of Science and Technology of China, Hefei, 230026, China \\
$^3$Department of Physics, University of California, San Diego, San Diego, California 92093, USA \\
$^4$ Institute of Physics, Chinese Academy of Sciences, Beijing 100080, Peoples Republic of China \\
$^5$ Collaborative Innovation Center of Quantum Matter, Beijing, China \\
$^6$Department of Physics and Astronomy, and Rice Center for Quantum Materials,
Rice University, Houston, TX 77251, USA \\
$^7$Center for Cold Atom Physics, Chinese Academy of Sciences, Wuhan 430071, P. R. China
}

\pacs{xxxxxx}
\begin{abstract}
\textbf{We present a flexible scheme to realize exact flat Landau levels on curved spherical geometry in a system of spinful cold atoms. This is achieved by Floquet engineering of a magnetic quadrupole field. We show that a synthetic monopole field in real space can be created. We prove that the system can be exactly mapped to the electron-monopole system on sphere, thus realizing Haldane's spherical geometry for fractional quantum Hall physics. The scheme works for either bosons or fermions. We investigate the ground state vortex pattern for an $s$-wave interacting atomic condensate by mapping this system to the classical Thompson's problem. We further study the distortion and stability of the vortex pattern when dipolar interaction is present. Our scheme is compatible with current experimental setup, and may serve as a promising route of investigating quantum Hall physics and exotic spinor vortex matter on curved space.}
\end{abstract}
\maketitle

%

{\em Introduction.}---
The realization of quantum Hall physics (QHP) in neutral atoms remains one of the long-standing goals in cold atom community \cite{pitaevskii2016bose,lewenstein2012ultracold,dalibard2011colloquium,jin2012introduction,galitski2013spin,zhou2013unconventional,zhai2012spin,jaksch2003creation,osterloh2005cold}. 
Theoretically, atomic systems not only provide an excellent platform to explore such novel physics for both bosons and fermions, the former is beyond the usual condensed matter systems, but also enable us to test various predictions with high precision due to its cleanness and high controllability. 
Experimentally, exact flat landau levels can be obtained in principle by rotating the confining harmonic potential \cite{viefers2008quantum,fetter2009rotating,cooper2008} with frequency equal to that of the harmonic trap. 
However, in this limit, effective trapping potential vanishes, and the atomic cloud loses confinement. This makes it almost impossible to reach the exact quantum Hall regime using this setup \cite{cooper2008}.
Therefore, searching for new flexible methods of realizing QHP becomes important.

On the other hand, QHP becomes more clear in a modified geometry, as pointed out by Haldane in 1983 \cite{haldane1983fractional} , who showed that a spherical surface trap with monopole \cite{shnir2006magnetic,dirac1931quantised,wu1976dirac} at the origin can be used as a prototype to understand such novel physics. The simplicity of this mode not only makes it an ideal numerical starting point to tackle this complex many-body system \cite{fano1986configuration,PhysRevLett.54.237}, but also reveals how interesting physics can be induced in curve spaces with the help of magnetic monopoles. Unfortunately, direct realization of this beautiful model seems impossible as no real magnetic monopole has been found.

In this paper, we show that, within current technique, exact Landau levels on Haldane's spherical geometry can indeed be implemented in a highly controllable manner. The key ingredient is the construction of synthetic monopole field in real space \footnote{We note that, the current synthetic magnetic monopoles found in a spinor condensate are defined by the order parameters of the system \cite{PhysRevLett.91.190402,pietila2009creation,ray2014observation,ray2015observation,Ollikainen2017,sugawa2016,ho2017}, which cannot be used to induce an effective magnetic force on neutral atoms, as required by QHP.}. We prove that this is possible by using atoms with internal spin degrees of freedom \cite{wu2012mott,ho1999pairing,yang2016topological,Kawaguchi2008,lahaye2009physics,kawaguchi2012spinor,griesmaier2005bose,aikawa2012bose,lu2011strongly} subjected to a Floquet engineered magnetic quadrupole field. By projecting the atom into the lowest-energy spin manifold, we confirm that the single-particle physics is mapped to an electron-monopole system \cite{haldane1983fractional,shnir2006magnetic} on sphere. This is exactly the Hamiltonian on curved sphere with flat Landau levels, as originally envisioned by Haldane, which enables the exploration of QHP using neutral atoms with high tunability.

As a first step, we investigate the exotic ground state vortex pattern in this curved geometry for Bose condensates.
For isotropic $s$-wave interaction, we show that stable vortex pattern can be well described by the standard Thompson's problem, which serves as a direct verification of charge-vortex duality in 2D system.
For dipolar atoms, the anisotropy of dipole-dipole interaction breaks the rotational symmetry, which thus results in the accumulation of vortices around the two poles and the equator, and can lead to an instability. 
We note that the effect of the underlying geometry on various quantum orders has been widely considered \cite{ho2015spinor,batle2016generalized,fomin2012tunable,PhysRevLett.114.197204,parente2011spin,Adhikari2012,li2015topological,imura2012spherical,kraus2008testing,moroz2016chiral,shi2015emergent,zhang2017potential},
and only addressed recently by Ho and Huang \cite{ho2015spinor} for condensates on a cylindrical surface.
Our work thus provides a promising route of exploring various novel spinor vortex matter involved in a curved  spherical geometry.

{\em Realization of synthetic monopole field.}---
Our scheme of realizing the synthetic monopole field for cold atoms can be outlined as follows.
We start by considering an atom with hyperfine spin ${\bf F}$ subject to a magnetic field
\bea
   \textbf{B}=B_0 \vec{z} + B_1 [1-4\lambda \cos (\omega t )](x\vec{x}+y\vec{y}-2z\vec{z}).
\eea
The magnetic field consists of a strong static bias field along the $z$-axis with magnitude $B_0$, and a time-periodic quadruple field with driving frequency $\omega$. The interaction between the atomic magnetic dipole and the field leads to the Zeeman Hamiltonian $\tilde{H}_{F}=-\mu_B g_F \textbf{F}\cdot \textbf{B}$
with $\mu_B$ the Bohr magneton and $g_F$ the corresponding Land\'{e} $g$-factor. For simplicity, here we neglect the quadratic Zeeman term proportional to $(\textbf{B} \cdot \textbf{F})^2 \simeq B_0^2 F_z^2$, which can be compensated by a proper choice of $\lambda$ (See Appendix A for details). 

The effects induced by the strong bias field $B_0$ can be removed by transforming the whole system into the rotating frame defined by the unitary operator
\begin{equation}
U = \exp(-i\omega_L t F_z),
\label{rotframeT}
\end{equation}
where $\omega_L \equiv \mu_B g_F B_0/\hbar$ is the Larmor frequency for the bias field.
The Hamiltonian in the rotating frame is given by $H_F = U^{\dag} \tilde{H}_{F} U -i U^{\dag} \partial_t U
$.
Under the condition $\omega=\omega_L$, i.e., the driving frequency of the quadruple field matches with the larmor frequency, and furthermore when $\omega$ is much larger than all other energy scales, the Hamiltonian in the rotating frame takes the following form \cite{isoshima2000creation} $
H_F \simeq \mu_B g_F B_1 \left[2\lambda \left(x F_x  +y F_y  \right) + 2 z F_z \right]$,
where the fast oscillating terms have been neglected.
The above Hamiltonian can be recast into the form
\bea
H_F = 2 \mu_B g_F B_1 \lambda r (\textbf{F} \cdot \vec{e}_r+\gamma \cos\theta F_z),
\label{HF}
\eea
where $r=\sqrt{x^2+y^2+z^2}$ is the radial coordinate, $\vec{e}_r$ the radial unit vector, and $\gamma \equiv 1/\lambda-1$. In the following, we will mainly focus on the situation $\lambda=1$ or $\gamma=0$, under which Hamiltonian (\ref{HF}) describes an atom with magnetic dipole moment moving in a radial magnetic field, whose strength increases linearly with $r$. If the atom is confined on a spherical shell surface (which, as will be shown below, will be the case we will focus on), this radial magnetic field is equivalent to a monopole field.

{\em Single-particle Hamiltonian.}---
Now we consider the full single-particle Hamiltonian which includes $H_F$ (with $\gamma=1$) and an isotropic harmonic trapping potential $V = m\omega_T^2 r^2/2$ with $m$ being the atomic mass and $\omega_T$ the harmonic trap frequency. In the unit system defined by $\hbar=m=\omega_T=1$, the single-particle Hamiltonian takes the form
\bea
{H_0}= -\frac{\vec{\nabla}^2}{2 } + \frac{1}{2} r^2 + \alpha' r \,\textbf{F} \cdot \vec{e}_r,  \label{singPH}
\eea
where $\alpha'\equiv 2 \mu_B g_F B_1 (\hbar m \omega^3_T)^{-1}$ measures the strength of Zeeman coupling with the synthetic monopole field.
Here and in the following, we assume $\alpha'>0$ without loss of generality. Under this convention, the lowest spin manifold corresponds to the spin state which is polarized along the local monopole field and obeys $\textbf{F} \cdot \vec{e}_r|F, -F\rangle_{\textbf{r}}= -F|F, -F\rangle_{\textbf{r}}$.
In the $F_z$-representation, we have $|F, -F\rangle_{\textbf{r}} =  \exp(-iF_z\varphi) \exp(-i F_y \theta) |F,-F \rangle_{\bf z}$, where $\theta$ and $\varphi$ are the polar and the azimuthal angles, respectively.

Under the assumption that the atom adiabatically follows the local monopole field and thus stays in the lowest spin manifold, we can write the total wave function of the atom as $\psi(\vec{r})=\phi(\vec{r}) |F, -F\rangle_{\textbf{n}}$, where $\phi(\vec{r})$ is the spatial wave function.
After projecting out the spin component, we find that $\phi(\vec{r})$ is governed by the following effective Hamiltonian (see Appendix B for detailed derivation)
\begin{equation}
H_{\rm eff} = \frac{(-i\vec{\nabla}+ \vec{A})^2}{2} + V(r),   \label{redH}
\end{equation}
where $\vec{A}(\vec{r}) = F \frac{\cos\theta}{r\sin\theta} \,\vec{e}_{\varphi}$
is the effective gauge potential, and $V(r)=r^2/2- \alpha r+F/2r^2$ with $\alpha=\alpha'F$ the effective trapping potential. When $\alpha \gg 1$, $V(r)$ has a minimum at $r =R \approx \alpha$, and the atom is tightly confined near this minimum with negligible radial excitation. Under this condition, the radial degrees of freedom is frozen and the spatial wave function is reduced to $\phi(\vec{r}) = f(\theta, \varphi)$, governed by the reduced Hamiltonian
\bea
H=\frac{1}{2R^2}\,\Lambda^2
\label{RH1}
\eea
with ${\boldsymbol \Lambda} = \vec{r}\times[-i \vec{\nabla} + \vec{A}(\vec{r})] $.

Hamiltonian (\ref{RH1}) describes a charged particle confined on a spherical surface with radius $R$ subject to a magnetic monopole with charge proportional to $F$ centered at the origin. The single-particle physics was studied by Dirac, Wu-Yang, and many others to clarify the quantization of monopole charge \cite{dirac1931quantised,wu1976dirac}, and later used by Haldane as an alternative spherical geometry to understand the Fractional QHP (FQHP) \cite{haldane1983fractional}. The single-particle eigenstates are given by the monopole harmonics $\mathcal{Y}^m_{l,F}$ \cite{greiter2011landau,fakhri2007new}
with $l=F,F+1,\cdots$ and $m=-l,-l+1,\cdots,l$.
The corresponding energy eigenvalues are  \[ E_k=[l(l+1)-F^2]/(2R^2) \]  with $k=l-F$, which leads to a Landau-level like structure.

The above construction of Haldane spherical surface provides a unique way to explore FQHP using neutral atoms. First, unconstrained expansion of the atomic cloud is avoided due to the finite size of the surface. The exact flatness of Landau levels enables that FQHP can be easily manifested with only a few particles by tunning the interaction effect via Feshbach resonance using magnetic field, or by changing the density of the cloud \cite{viefers2008quantum,fetter2009rotating,cooper2008}. Second, the simplicity of the model makes it possible for the direct comparison between experimental and theoretical results \cite{fano1986configuration,PhysRevLett.54.237}, which provides an ideal testbed for various theoretical predictions about FQHP. Third, the high flexibility of the system also enables the investigation of novel quantum matter related to QHP and curved spherical geometry, as we will show in the following.

{\em Ground state vortex structure for condensate with contact interaction.}---
We now consider the properties of a weakly-interacting atomic condensate. The reduced condensate wave function $f (\hat{\Omega})=f(\theta , \varphi)$ satisfies the following Gross-Pitaevskii (GP) equation:
\bea
i \partial_t f(\hat{\Omega})  = \left[ \frac{\Lambda^2}{2R^2} + g |f(\hat{\Omega})|^2 + g_d D(\hat{\Omega}) \right] f(\hat{\Omega}),
\eea
where the second term in the square bracket describes the contact $s$-wave interaction characterized by the dimensionless interaction strength $g \simeq 2\sqrt{2 \pi}N a /(l_T R^2)$, with $a$ the $s$-wave scattering length and $N$ the number of atoms; the third term in the square bracket describes the dipolar interaction characterized by strength $g_d$, and the form of $D(\hat{\Omega})$ depends on the orientation of the atomic dipole, whose explicit form is given in Appendix D.

For non-dipolar condensate with $g_d=0$, since the single-particle eigenstates in the lowest Landau level can be written as $\mathcal{Y}^m_{F,F} \sim u^{F-m}v^{F+m}$ with $u=\cos\frac{\theta}{2}e^{-i\frac{\varphi}{2}}$ and $v=\sin\frac{\theta}{2}e^{i\frac{\varphi}{2}}$, a general wave function within this subspace can be expressed as $f(\hat{\Omega})=\sum_{m=0}^{2F}c_m u^{F-m}v^{F+m}$, which can then be factorized as $\Pi_{j=1}^{2F}(uv_j-u_jv)$ up to a normalization constant. This describes a lattice of $2F$ vortices, with $(u_j,v_j)$ representing the coordinates of the vortices on the sphere. The interaction energy is written as $U_{\rm int} \sim \int d \hat{\Omega} |f(\hat{\Omega})|^4$ with
\bea
|f(\hat{\Omega}|^2= e^{-K},\mbox{ and }\vspace{.7cm} K \sim -2\sum_j\ln|uv_j-u_jv|,
\eea
where $|uv_j-u_jv|$ is the chord distance between two points on the unit sphere. The quantity $K$ precisely describes the energy of an electron interacting with $2F$ other electrons located at $(u_j,v_j)$.
Therefore, minimization of $U_{\rm int}$ with respect to $(u_j,v_j)$ can be mapped to the problem of finding the stable configuration of $2F$ electrons on sphere. This is exactly the well-known Thomson's problem, as J. J. Thomson posed such a model to understand his plum pudding model of the atom in 1904 \cite{thomson1904xxiv}.

\begin{figure}[htpb]
  \centering
  \includegraphics[width=.90\linewidth]{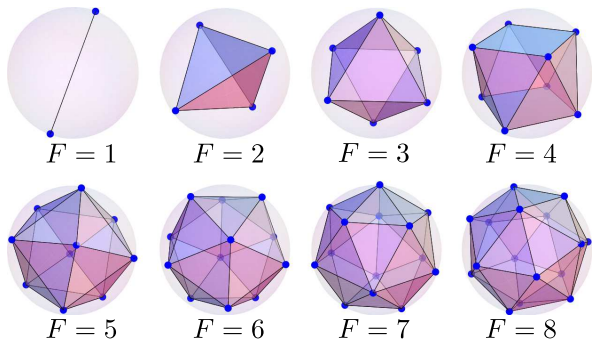}
  \caption{(Color) Stable ground-state vortex configurations with isotropic s-wave interaction $gR^2=50$ for different $F$. Here the locations of the vortex cores are represented with blue dots on the spherical surface.   }
  \label{fig1:thomson}
\end{figure}

Figure \ref{fig1:thomson} shows the configuration of the $2F$ vortices on the sphere obtained from our numerics for $F$ up to 8 by solving the GP equation using imaginary evolution method \cite{healy2003ffts,fftaddress}.
For each $F$, the pattern is equivalent to (up to a global rotation) the standard solutions to Thomson's problem, which is a direct reflection of charge-vortex duality in 2D systems.
We note that the singularity of $\vec{A}$ at both the south and the north poles in our chosen gauge has no effect after projecting the effective wave function back to the usual Zeeman manifolds.
For each Zeeman sublevel, there is a giant vortex with  $\mp F+F_z$ units of circulation  around the north and the south poles, respectively, as shown in Appendix D.

\begin{figure}[htpb]
  \centering
  \includegraphics[width=.9\linewidth]{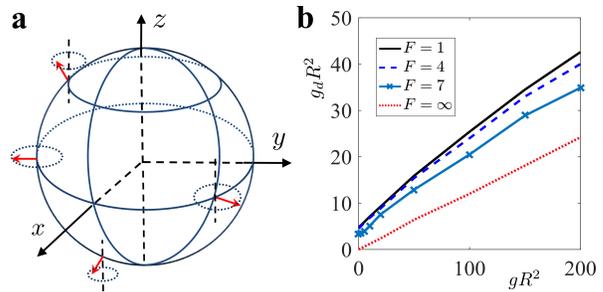}
  \caption{(Color) Modulated dipolar orientation on the spherical surface and the critical $g^{(c)}_d$ along with repulsion $g$ for different $F$. (a) In the interaction picture, each local dipole rotates around the $z$-axis with frequency $\omega_L$, which results in an average dipolar interaction described by Eq. \ref{reducedUd}. (b) shows the stability diagram in the $g_d-g$ plane. The condensates is unstable when $g_d>g^{(c)}_d$. The critical  $g^{(c)}_d$ decreases along with $F$, and approaches the limit dotted line for $F \rightarrow \infty$.}
  \label{criticalg}
\end{figure}

{\em Stability and vortex structure in dipolar condensate. }---
We now include the dipolar interaction term in the Hamiltonian, which is very typical for condensates of atoms with large internal spin. For an atom with spin ${\bf F}$, it possesses a magnetic dipole moment $\vec{\mu}=\mu_B g_F {\bf F}$. Given two dipoles $\vec{\mu}_1$ and $\vec{\mu}_2$ located at $\vec{r}_1$ and $\vec{r}_2$, respectively,
the dipolar interaction in the lab frame reads
$
U_{d} (\vec{r}_1,\vec{r}_2) =  [\vec{\mu}_1\cdot\vec{\mu}_2-3(\hat{r}_{12}\cdot\vec{\mu}_1)
(\hat{r}_{12}\cdot\vec{\mu}_2)]/r_{12}^3
$
with ${r}_{12}=|\vec{r}_1-\vec{r}_2|$ and $\hat{r}_{12} =(\vec{r}_1-\vec{r}_2)/r_{12}$. In the rotating frame defined by the unitary operator $U$ in Eq.~(\ref{rotframeT}), it transforms as $U_{d} (\vec{r}_1,\vec{r}_2) \rightarrow U^{\dag}U_{d}U$, and becomes time-dependent. In this case, each local spin rotates around the $z$-axis with the frequency $\omega_L$, as shown in Fig.~\ref{criticalg}a. After integrating out the high-frequency parts, we arrive at an effective time-independent dipolar interaction potential as (see Appendix E for details)
\bea
U_d^{(e)}(\vec{r}_1,\vec{r}_2)=\frac{1}{{r}_{12}^3} \sqrt{\frac{6\pi}{5}}Y^0_2(\hat{\Omega}_{12})\Sigma^0_2(\vec{\mu}_1,\vec{\mu}_2) \label{reducedUd}
\eea
with $\hat{\Omega}_{12}$ the orientation of $\vec{r}_{12}$ and $\Sigma^0_2(\vec{\mu}_1,\vec{\mu}_2) = \sqrt{2/3}(\vec{\mu}_1\cdot \vec{\mu}_2 - 3 \mu_{1,z} \mu_{2,z})$.
Therefore, the interaction breaks the rotational symmetry, which modifies the distribution of the condensate and distorts the vortex pattern obtained in the previous section.

The anisotropicity of $U_d^{(e)}(\vec{r}_1,\vec{r}_2)$ can be illustrated from its local properties. For two neighboring sites represented as $\vec{r}'=\vec{r}+\delta (\cos\alpha \vec{e}_{\theta}+\sin\alpha \vec{e}_{\phi})$ with $\delta$ an infinitesimal arc length and $\alpha$ the azimuthal angle in the local tangent plane, the diploar interaction can be written as
\bea
U_d^{(e)}(\vec{r},\vec{r}') \propto \frac{(3\cos^2\alpha\sin^2\theta-1)(1-3\cos^2\theta)}{\delta^3}.  
\eea
When $\vec{r}-\vec{r}'$ is parallel with the longitude with $\alpha=0$, $U_d^{(e)}(\vec{r},\vec{r}')$ becomes attractive for $\theta \in [\theta_2, \theta_1]\cup [\pi-\theta_1, \pi-\theta_2]$ with $\theta_1=\cos^{-1}\sqrt{1/3}$ and $\theta_2=\cos^{-1}\sqrt{2/3}$. However, when $\vec{r}-\vec{r}'$  coincides with the latitude, $U_d^{(e)}(\vec{r},\vec{r}')$ is attractive only around the equator with $\theta \in [\theta_1,\pi-\theta_1]$.  This azimuth-dependent attractive interaction can result in instability and collapses the condensates. The average local dipolar interaction can be estimated by integrating over the angle $\alpha$ and reads $\overline{U}_d^{(e)}(\vec{r},\delta) \propto (1-3\cos^2\theta)^2/(2\delta^3)$, which is minimized at $\theta=\theta_1$ and reaches its local maxima when $\theta=0$ (or $\pi$) and $\pi/2$. 

Figure \ref{criticalg}b shows the critical dipolar interaction strength $g^{(c)}_d$ as a function of contact interaction stregnth $g$ for different spin $F$.
The condensate is stable below the critical line, and unstable above it.
The critical $g^{(c)}_d$ deceases as $F$ increases.
Physically, this can be understood by noticing that the degeneracy of single-particle ground state (i.e., the lowest Landau level) increases linearly with $F$.
For larger $F$, the condensate has more degrees of freedom to adjust its wave function within the lowest Landau level to lower the interaction energy, while the kinetic energy almost remains unaffected.
In the limiting case $F \rightarrow \infty$, the kinetic energy is complete quenched, and the stability of the condensate is solely determined by the relative strength of the contact repulsion and the dipolar interaction. This critical $g_d^{(c)}$ for $F=\infty$ is represented by the dotted line shown in Fig. \ref{criticalg}. Our numerical results for finite $F$ provides a direct verification towards this limit.

\begin{figure}[htpb]
  \centering
  \includegraphics[width=.90\linewidth]{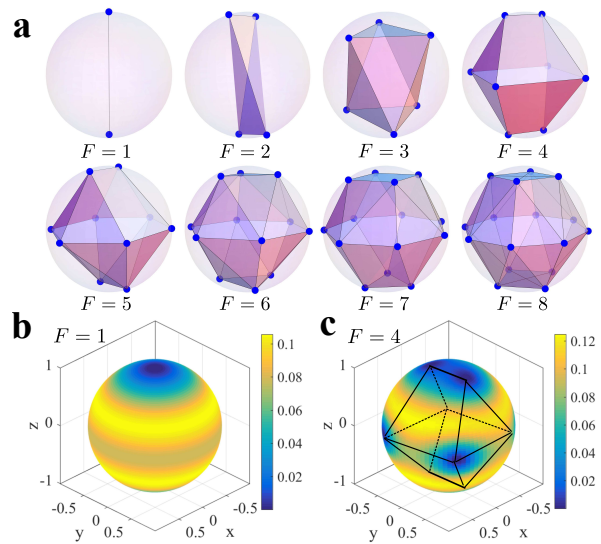}
  \caption{(Color) Stable ground-state vortex configurations of dipolar BEC with $gR^2=50$ and different $g_dR^2$. In (a), the data is obtained using $g_dR^2=10$ for $F=1\sim 5$ and $g_dR^2=8$ for $F=6\sim8$ respectively. The vortex cores are represented with blue dots on the spherical surface.  Since the dipole-dipole interaction breaks the rotation symmetry, the pattern is only equivalent up to a global rotation around $z$-axis for each $F$. (b) and (c) show the selected density portraits for $F=1$ and $F=4$, respectively. The vortices pattern is guided by lines and dot lines inside the sphere.  }
  \label{thomson-dipolar}
\end{figure}

We note that for condensates with stronge dipolar interaction near the critical $g^{(c)}_d$, the vortex configuration also deviates significantly from the standard Thomson's lattice, particularly for large $F$, as depicted in Fig. \ref{thomson-dipolar}. Compared with $s$-wave contact interaction, these patterns are equivalent up to a global rotation around the $z$-axis. Since $\overline{U}_d^{(e)}(\vec{r},\delta)$ reaches its local maxima at $\theta=0$ ($\pi$)and $\pi/2$, to minimize the interaction energy, vortices appear first near the two poles, and later spread around the equator with the increasing of $F$. For all $F$, the density peaks around the two latitude lines with $\theta \sim \theta_1$ and $\pi-\theta_1$ respectively, as shown in Fig. (3b)-(3c). This can be understood from the average local dipolar interaction as $\overline{U}_d^{(e)}(\vec{r},\delta)$ is minimized when $\theta=\theta_1$ and $\pi-\theta_1$.

{\em Experimental feasibility.}---
For $^{168}$Er \cite{aikawa2012bose} atoms with $F=6$ in a bias magnetic field $\vec{B}=B_0 \vec{z}$ with $B_0=0.30G$, the linear Zeeman splitting is about $\omega=2\pi\times 500 kHz$, which is much larger than the typical trapping frequency $\omega_T=2\pi\times 2.5 kHz$.
The radial length scale of the condensates is estimated as $l_T=\sqrt{\hbar/(m\omega_T)}\simeq 0.246 \mu m$. To obtain an effective spherical-shell trap, we need $\alpha=2\mu_B g_Fl_TB_1F/(\hbar\omega_T) \simeq 0.48 B_1[cm\cdot G^{-1}] \gg 1$.
For current experimental setup, it is not difficult to achieve a magnetic field gradient of $B_1 \simeq 20 G\cdot cm^{-1}$.
This leads to $\alpha  \sim R \sim 10 \gg 1 $.
Using these setting, the splitting between different Landau levels is  $\Delta_L = (F+1)\omega_T/R^2 \sim 2\pi \times 175 Hz$, which is much smaller compared with the excitation energy $\omega_T$ along the radial direction.
In this case, the condensates are confined only near the surface of the sphere, as required by our derivation. We note that for $^{87}$Rb atoms with $F=1$ \cite{xu2013atomic,anderson2013magnetically,luo1502tunable}, $B_1$ needs to be as high as $10^2 \sim 10^3 G\cdot cm^{-1}$, which is still a challenge to current experimental setup.
Therefore, atomic condensates with large internal spin are always helpful for the construction of such surface trap.
The contact interaction can be estimated by $gR^2=2\sqrt{2\pi} aN/l_T\simeq 0.236 N$, which can then be tuned over a wide parameter regimes.

{\em Conclusion.}--- 
By constructing an effective hedgehog-like gradient magnetic field with spinful atoms, we have proposed a flexible way to implement an effective electron-monopole system confined on a spherical surface. We show how various vortex patterns can be obtained in the presence of inter-particle interactions.
The scheme proposed here provides a promising route to investigate FQHP of bosons or fermions in curved space. Finally, the synthetic hedgehog-like gradient magnetic field for spinful atoms also provides new possibilities of searching for exotic spinor quantum matters related to magnetic monopoles \cite{PhysRevLett.91.190402,pietila2009creation,ray2014observation,ray2015observation,Ollikainen2017,Kawaguchi2008}.

\vspace{.5cm}\noindent
\textbf{Acknowledgements} \newline
This work was funded by National Natural Science Foundation of China (Grant Nos.11474266,11574294,11774332), the Major Research plan of the NSFC (Grant No. 91536219), the National Plan on Key Basic Research and Development (Grant No. 2016YFA0301700), and the ``Strategic Priority Research Program(B)" of the Chinese Academy of Sciences (Grant No. XDB01030200). CW is supported by the NSF DMR-1410375 and AFOSR FA9550-14-1-0168. HP is supported by the US NSF (Grant No. PHY-1505590) and the Welch Foundation (Grant No. C-1669). RW is supported by the National Basic Research Program of China (Grant No. 2013CB922002), and NSFC (No. 11474347).

\appendix

\section{Synthetic monopole fields for spinful atoms}
Let us start with the Zeeman Hamiltonian 
\bea
\tilde{H}_{F}=-\mu_B g_F \textbf{F}\cdot \textbf{B}
\eea 
for spinor atoms with hyperfine spin ${\bf F}$ subjecting to a time-dependent quadrupole magnetic field 
\bea
   \textbf{B}=B_0 \vec{z} + B_1 [1-4\lambda \cos (\omega t +\phi )](x\vec{x}+y\vec{y}-2z\vec{z}).
\eea
In the rotating frame defined by the unitary operator
\bea
U = \exp(-i\omega_L t F_z),
\label{rotframe}
\eea
the Hamiltonian is given by 
\bea
H_F = U^{\dag} \tilde{H}_{F} U -i U^{\dag} \partial_t U.
\eea
When $\omega=\omega_L$, using the following identities
\bea
U^{\dag} F_x U = F_x \cos(\omega t) - F_y \sin(\omega t), \nn \\
U^{\dag} F_y U = F_y \cos(\omega t) + F_x \sin(\omega t), \nn
\eea
we have
\bea
H_F &=& -\mu_B g_F [1-4\lambda \cos (\omega t +\phi )] [F_x(x\cos\omega t+y\sin\omega t) \nn \\
&&+F_y(-x\sin(\omega  t)+y\cos\omega t)-2zF_z].
\eea
Since
\bea
\cos(\omega t +\phi)\cos\omega t &=& \frac{1}{2}[\cos(2\omega t +\phi)+\cos\phi], \nn \\
\cos(\omega t +\phi)\sin\omega t &=& \frac{1}{2}[\sin(2\omega t +\phi)-\cos\phi],
\eea
after a simple algebra, we arrive 
\bea
H_F &=& -\mu_B g_F \big[-2\lambda F_x(x\cos\phi-y\sin\phi) \nn \\
&& \;\;\;\;\;\;\;\;\;\;\;\;\;\; -2\lambda F_y(-x\sin\phi+y\cos\phi) \nn \\
&& \;\;\;\;\;\;\;\;\;\;\;\;\;\; -2zF_z  \big] +H'_F(t)
\eea
with 
\bea
H'_F(t)&=&-\mu_B g_F \big\{ F_x(x\cos\omega t+y\sin\omega t) \nn \\
&&+F_y(-x\cos\omega t + y\cos\omega t)-8\lambda z F_z\cos(\omega t+\phi)  \nn \\
&& - 2 \lambda F_x [x\cos(2\omega t+\phi)+y\sin(2\omega t+\phi)] \nn \\
&& - 2 \lambda F_x [-x\sin(2\omega t+\phi)+y\cos(2\omega t+\phi)] \big \}. \nn \\
\eea
When $\omega$ is much larger than all other energy scales, $H'_F(t)$ can be safely neglected. 
By setting $\lambda=1$ and $\phi=0$, we obtained the desired hedge-hog like effective magnetic field shown in the main text.

\section{reduced dynamics on spherical surface}
For effective Zeeman term proportional to $\alpha r \textbf{F}\cdot \vec{e}_r$, the lowest Zeeman sublevel is orientation-dependent and reads
\bea
\textbf{F} \cdot \vec{e}_r(\theta,\varphi) |F,- F\rangle_{\textbf{r}}=- F|F,- F\rangle_{\textbf{r}}
\eea
with the azimuth angle defined as $\textbf{r}(r,\theta,\varphi)=r(\sin\theta\cos\varphi,\sin\theta \sin \varphi, \cos \theta)$, and $F$ the hyperfine spin of the atomic species. For latter purpose, it is much convenient to
rewrite them using the common eigenvectors defined by $\{ \textbf{F}^2, F_z \}$. In this basis, we have \cite{isoshima2000creation,ho2015spinor}
\bea
|\overline{F}\rangle_{\textbf{r}}\equiv|F,- F\rangle_{\textbf{r}} = S^{\dag}|F,- F\rangle_{\bf z}
\eea
with $S^{\dag} = \exp(-iF_z\varphi) \exp(-i F_y \theta)$.
Here the angular momentum operators are defined as
\bea
(F_x)_{mn}&=&\frac{1}{2} [\sqrt{(2F+2-m)(m-1)} \delta_{m-1,n} \nn \\
&&+ \sqrt{(2F+1-m)m} \delta_{m+1,n}], \\
(F_y)_{mn}&=&\frac{1}{2} [i\sqrt{(2F+2-m)(m-1)} \delta_{m-1,n} \nn \\
&&-i\sqrt{(2F+1-m)m} \delta_{m+1,n}]                           , \\
(F_z)_{mn}&=&\delta_{mn}(F+1-m)
\eea
with
\bea
(m,n)=(1,2,\cdots,2F+1),
\eea
and satisfy the commutation relation $[F_{\alpha},F_{\beta}]=i\epsilon_{\alpha\beta\gamma}F_{\gamma}$. Using the following identities
\bea
e^{-i\theta F_y} F_z e^{i \theta F_y} = F_z \cos\theta + F_x \sin\theta, \nn \\
e^{-i\varphi F_z} F_x e^{i \varphi F_z} = F_x \cos\varphi + F_y \sin\varphi, \nn
\eea
we have
\bea
\textbf{F} \cdot \vec{e}_r (\theta,\varphi) &=& \sin\theta (\cos\varphi F_x + \sin\varphi F_y) +\cos\theta F_z \nn \\
&=& e^{-i\varphi F_z}e^{-i\theta F_y} F_z e^{i\theta F_y} e^{i\varphi F_z} \nn \\
&=& S^{\dag} F_z S.
\eea

When the local Zeeman fields is sufficiently strong, the spin is polarized by the effective local magnetic field and the wave function of the atom can then be approximated as  $\psi(\vec{r})=\phi(\vec{r})|\overline{F}\rangle_{\textbf{r}}$. The kinetic energy can then be estimated as $E_{int}=\int d\vec{r} \vec{\nabla}\psi^{\dag} \cdot \vec{\nabla}\psi$ with
\bea
\vec{\nabla}\psi^{\dag} \cdot \vec{\nabla}\psi &=& \Big[ \vec{\nabla} \phi^* + {}_{\textbf{r}}\langle \vec{\nabla}\overline{F}|\overline{F} \rangle_{\textbf{r}} \phi^*\Big]\cdot \Big[\vec{\nabla} \phi + {}_{\textbf{r}}\langle \overline{F}|\vec{\nabla}\overline{F} \rangle_{\textbf{r}} \phi \Big]  \nn \\
&& + \Big[{}_{\textbf{r}}\langle \vec{\nabla}\overline{F}|\vec{\nabla}\overline{F} \rangle_{\textbf{r}} - {}_{\textbf{r}}\langle \vec{\nabla}\overline{F}|\overline{F} \rangle_{\textbf{r}} {}_{\textbf{r}}\langle \overline{F}|\vec{\nabla}\overline{F} \rangle_{\textbf{r}} \Big] |\phi|^2. \nn
\eea
To obtain the reduced dynamics within this subspace, we need to calculate
\bea
{}_{\textbf{r}}\langle \overline{F}|\vec{\nabla}\overline{F} \rangle_{\textbf{r}} &=& {}_{\textbf{z}}\langle \overline{F}|S\cdot\vec{\nabla}S^{\dag}|\overline{F} \rangle_{\textbf{z}}  \nn \\
&=& {}_{\textbf{z}}\langle \overline{F}|e^{i\theta F_y} (-iF_z \vec{\nabla} \varphi -iF_y \vec{\nabla} \theta) e^{-i\theta F_y}|\overline{F} \rangle_{\textbf{z}} \nn \\
&=& i F\cos\theta \vec{\nabla} \varphi  \nn   \\
{}_{\textbf{r}}\langle \vec{\nabla} \overline{F}|\vec{\nabla}\overline{F} \rangle_{\textbf{r}} &=& {}_{\textbf{z}}\langle \overline{F}|\vec{\nabla}S\cdot\vec{\nabla}S^{\dag}|\overline{F} \rangle_{\textbf{z}}  \nn \\
&=& {}_{\textbf{z}}\langle \overline{F}|e^{i\theta F_y} (F_z \vec{\nabla} \varphi +F_y \vec{\nabla} \theta)^2 e^{-i\theta F_y}|\overline{F} \rangle_{\textbf{z}} \nn \\
&=& (F^2\cos^2\theta+\frac{F}{2}\sin^2\theta)|\vec{\nabla} \varphi|^2 +\frac{ F}{2} |\vec{\nabla}\theta|^2, \nn
\eea
where we have used the relations
\bea
{}_{\textbf{z}}\langle \overline{F}|F_x^2|\overline{F} \rangle_{\textbf{z}} &=& {}_{\textbf{z}}\langle \overline{F}|F_y^2|\overline{F} \rangle_{\textbf{z}}=\frac{F}{2}, \nn \\
{}_{\textbf{z}}\langle \overline{F}|F_x|\overline{F} \rangle_{\textbf{z}} &=& {}_{\textbf{z}}\langle \overline{F}|F_y|\overline{F} \rangle_{\textbf{z}}=0 \nn \\
\vec{\nabla} \varphi \cdot \vec{\nabla} \theta &=& 0 \nn.
\eea
Using $\vec{\nabla} \varphi = \vec{e}_{\varphi}/(r\sin\theta)$ and $\vec{\nabla} \theta = \vec{e}_{\theta}/r$, the effective potential induced by the spin wave function is given by
\bea
&& {}_{\textbf{r}}\langle \vec{\nabla}\overline{F}|\vec{\nabla}\overline{F} \rangle_{\textbf{r}} - {}_{\textbf{r}}\langle \vec{\nabla}\overline{F}|\overline{F} \rangle_{\textbf{r}} {}_{\textbf{r}}\langle \overline{F}|\vec{\nabla}\overline{F} \rangle_{\textbf{r}} \nn \\
&=& \frac{F}{2}\sin^2\theta|\vec{\nabla} \varphi|^2 +\frac{ F}{2} |\vec{\nabla}\theta|^2 =\frac{ F}{r^2},
\eea
which is orientation-independent due to the center-symmetry of the system.

Using the above formula, we can rewrite
\bea
\vec{\nabla}\psi^{\dag} \cdot \vec{\nabla}\psi &=& \Big[{}_{\textbf{r}}\langle \overline{F}|\vec{\nabla}\phi^* + {}_{\textbf{r}}\langle \vec{\nabla}\overline{F}|\phi^*\Big]\cdot\Big[\vec{\nabla}\phi |\overline{F}\rangle_{\textbf{r}} + \phi \vec{\nabla} |\overline{F}\rangle_{\textbf{r}} \Big] \nn \\
&=& \Big[ \vec{\nabla} \phi^* + {}_{\textbf{r}}\langle \vec{\nabla}\overline{F}|\overline{F} \rangle_{\textbf{r}} \phi^*\Big]\cdot \Big[\vec{\nabla} \phi + {}_{\textbf{r}}\langle \overline{F}|\vec{\nabla}\overline{F} \rangle_{\textbf{r}} \phi \Big]  \nn \\
&& + \Big[{}_{\textbf{r}}\langle \vec{\nabla}\overline{F}|\vec{\nabla}\overline{F} \rangle_{\textbf{r}} - {}_{\textbf{r}}\langle \vec{\nabla}\overline{F}|\overline{F} \rangle_{\textbf{r}} {}_{\textbf{r}}\langle \overline{F}|\vec{\nabla}\overline{F} \rangle_{\textbf{r}} \Big] |\phi|^2, \nn \\
&=& |(-i\vec{\nabla}'+\vec{A})\phi|^2 + \frac{F}{r^2} |\phi|^2
\eea
with $\vec{A}= F \cos\theta \vec{e}_{\varphi}/(r\sin\theta)$. The total energy functional can then be simplified as
\bea
\frac{E_0[\psi]}{\hbar\omega} &=& \int d\vec{r} \Big [\frac{1}{2}|\vec{\nabla}\psi|^2 + (\frac{1}{2}r^2 - \alpha' F r) |\psi|^2]  \nn \\
&=&  N \int d\vec{r} \Big [\frac{1}{2}|(-i\vec{\nabla}+\vec{A})\overline{\phi}|^2 + V(r) |\overline{\phi}|^2],
\eea
where $N$ is the total number of particles, $\overline{\phi}=\phi/\sqrt{N}$ is the normalized wave function, and $V(r)$ is the total effective central potential
\bea
V(r)=\frac{1}{2}r^2-\alpha r + \frac{F}{2r^2}
\eea
with $\alpha = \alpha' F$. For large $\alpha \gg 1$, the condensates are mainly trapped near the surface a sphere with radius $R\sim \alpha$. The radial part of the condensates wavefunction can be approximated by $h(r) \simeq  (\pi r^2)^{-1/4} \exp[-(r-\alpha)^2/2]$ with total wavefunction $\overline{\phi}(\vec{r}) = h(r)f(\hat{\Omega})$. The kinetic energy can then be simplified as
\bea
|\hat{P}\overline{\phi}|^2 &=& |(-i\vec{\nabla}+\vec{A})\overline{\phi}|^2 = |\hat{P}_r\overline{\phi}|^2 + |\hat{P}_{\Omega}\overline{\phi}|^2
\eea
with
\bea
\hat{P}_r &=& -i\partial_r, \nn \\
\hat{P}_{\Omega}&=&(-i)[\vec{e}_{\theta}\partial_{\theta}+\vec{e}_{\varphi}\sin^{-1} \theta (\partial_{\varphi} + i F\cos\theta)]. \nn
\eea
Based on this approximation, the reduced dynamics shown in Eq. (8) can be obtained accordingly.

The presence of non-zero $\gamma$ introduces an additional term proportional to $\alpha'\gamma r\cos\theta F_z$ into the single-particle Hamiltonian. When $\gamma \ll 1$, The condensates are mainly trapped near the surface with the radius $r \sim R$, which is the case we focus on. This term induces an effective surface trapping potential in the spin manifold approximated by
\bea
V(\theta)&=&\alpha' \gamma r \cos\theta {}_{\textbf{r}}\langle \overline{F}| F_z |\overline{F}\rangle_{\textbf{r}} \nn \\
&=&\alpha'\gamma r \cos\theta {}_{\textbf{z}}\langle \overline{F}| S F_z S^{\dag} |\overline{F}\rangle_{\textbf{z}}  \nn \\
 &\simeq & -\gamma \alpha R \cos^2\theta  \label{vtheta}
\eea

There also exist other potential experimental imperfections which may deform  the spherical trap. First, the gravitation potential breaks the rotational symmetry, but it can be readily compensated by an additional magnetic gradient as routinely done in labs. Secondly, due to the quadratic Zeeman effect, the presence of large constant bias field leads to another energy shift proportional to $(\textbf{B} \cdot \textbf{F})^2 \simeq B_0^2 F_z^2$. After projecting into the manifold defined by the spherical shell, this energy shift gives additional effective spherical trap potential as $V'(\theta) \simeq B_0^2 F(F-\frac{1}{2}) \cos\theta^2 $. This imperfection can be, however, cancelled with a proper choice of finite $\gamma$, such that the resulting additional potential $V(\theta)$ [see Eq.~(\ref{vtheta})] can compensate $V'(\theta)$. Therefore, perfect symmetric trap can be prepared within current experimental setup.

\section{ladder operators for monopole harmonics}

Given the single-particle Hamiltonian
\bea
H=\frac{1}{2R^2}\,\Lambda^2
\label{HR}
\eea
with ${\boldsymbol \Lambda} = \vec{r}\times[-i \vec{\nabla} + \vec{A}(\vec{r})] $,
we can define the angular operator ${\bf L}={\boldsymbol \Lambda} -F\vec{e}_r$ which satisfies the commutation relation 
\[ [L_{\alpha},X_{\beta}]=i \epsilon_{\alpha\beta\gamma}X_{\gamma} \]
with 
\[ X=\Lambda, L, \mbox{  and  } \vec{r}. \] 
Since ${\boldsymbol \Lambda} \cdot \vec{e}_r=\vec{e}_r \cdot {\boldsymbol \Lambda} =0$, we have $L^2=\Lambda^2+F^2$. The single-particle eigenstates can be constructed as the common eigenvectors of $L^2$, $\Lambda^2$, and $L_z$, whose wave function are given by the monopole harmonics $\mathcal{Y}^m_{l,F}$ \cite{greiter2011landau,fakhri2007new} with 
\[ l=F,F+1,\cdots \mbox{ and } m=-l,-l+1,\cdots,l. \]
The corresponding energy eigenvalues are  
\[ E_k=[l(l+1)-F^2]/(2R^2) \] 
with $k=l-F$, which leads to a Landau-level like structure.

The nice analytical properties of $\mathcal{Y}^m_{l,F}$ allow us to define the relevant ladder operators \cite{greiter2011landau,fakhri2007new}, with which we can raise and lower the indices $l$, $m$, and $F$ respectively.
There are three different indices in $\mathcal{Y}^m_{l,F}$ which greatly enriches the internal structure of these monopole harmonics. Fortunately, we can define series ladder operators which enable us to raise and low these indices. For instance, the ladder operator for index $m$ can be defined as
\bea
L_+(m) \mathcal{Y}^m_{l,F} &=& \sqrt{(l-m)(l+m+1)} \mathcal{Y}^{m+1}_{l,F}, \nn \\
L_-(m) \mathcal{Y}^m_{l,F} &=& \sqrt{(l-m+1)(l+m)} \mathcal{Y}^{m-1}_{l,F}, \nn
\eea
where
\bea
L_+(m) &=& e^{i\varphi}\big[ \partial_{\theta}+i\frac{\cos\theta}{\sin\theta}\partial_{\varphi}-\frac{F}{\sin\theta}  \big], \nn \\
L_-(m) &=& e^{-i\varphi}\big[ -\partial_{\theta}+i\frac{\cos\theta}{\sin\theta}\partial_{\varphi}-\frac{F}{\sin\theta}  \big]. \nn
\eea
Similarly, monopole harmonics with different charge $F$ can be connected through
\bea
L_+(F) \mathcal{Y}^m_{l,F} &=& \sqrt{(l-F)(l+F+1)} \mathcal{Y}^m_{l,F+1}, \nn \\
L_-(F) \mathcal{Y}^m_{l,F} &=& \sqrt{(l-F+1)(l+F)} \mathcal{Y}^m_{l,F-1}, \nn
\eea
with
\bea
L_+(F) &=& \partial_{\theta} - \frac{m+F \cos\theta}{\sin\theta}, \nn \\
L_-(F) &=& -\partial_{\theta} - \frac{m+F \cos\theta}{\sin\theta}. \nn
\eea
Finally, the index of the total angular momentum $L$ can also be changed by
\bea
L_+(l) \mathcal{Y}^m_{l,F} &=& \sqrt{\frac{[(l+1)^2-m^2][(l+1)^2-F^2](2l+1)}{(l+1)^2(2l+3)}} \mathcal{Y}^m_{l+1,F}, \nn \\
L_-(l) \mathcal{Y}^m_{l,F} &=& \sqrt{\frac{[l^2-m^2][l^2-F^2](2l+1)}{l^2(2l-1)}} \mathcal{Y}^m_{l-1,F}, \nn
\eea
with
\bea
L_+(l) &=& \sin\theta \partial_{\theta}+(l+1)\cos\theta + \frac{mF}{l+1}, \nn \\
L_-(l) &=& -\sin\theta \partial_{\theta}+l\cos\theta + \frac{mF}{l}. \nn
\eea
Using these ladder operators, all monopole harmonics can be constructed by starting with the trivial case $\mathcal{Y}^{-F}_{F,F}=\sqrt{\frac{2F+1}{4\pi}} (\cos \frac{\theta}{2})^{2F}e^{-iF\varphi}$.

The effective trapping potential on sphere can also be expressed as the combination of these ladder operators
\bea
\cos^2 \theta =\Big[ \frac{L_+(l)+L_-(l+1)}{l+1}-\frac{2mF}{(l+1)^2} \Big]^2,
\eea
which couples different Landau levels with the same angular momentum number $m$, and can be used as a knob to tune the splitting of different levels.


\section{Thomson lattice on spherical surface for s-wave interaction}
For Bose condensates with s-wave interaction, the vortex pattern can be characterized by Thomson's problem, as shown in Fig. 1 in the main text. For instance, when $F=2$, $3$, and $6$, the optimal solutions are given by the well-known Platonic solids where each face corresponds to an equilateral triangle.
However, the stable configuration for $F=4$ vortices is the square antiprism instead of the usual cube, which indicates that stable equilibrium does not necessarily mean perfect symmetry.
Except for $F=4$, stable vortex configurations always prefer triangular faces with distinct symmetry instead of square faces.

Figure \ref{fig:Q=3all}(a) and (f) show the calculated vortex patten for $F=3$. The vortices are arranged to form a regular octahedron.
Since $\vec{A}$ has singular singularity at both the south and the north poles in our chosen gauge, there are two artificial giant coreless vortex at the two poles.
Physically, the location of the vortex cores can be obtained through the gauge-invariant velocity field
defined as $\vec v =\nabla \phi -\vec{A}$.
In this case, the true vorticity is the difference between the winding of $\phi$ and the flux of $\vec{A}$.
This singularity can also be removed by projecting the effective wave function back to the usual Zeeman manifolds as shown in Fig. \ref{fig:Q=3all}(b-e) and (g-j).
For each Zeeman sublevel, there is a giant vortex with  $\mp F+F_z$ units of circulation  around the north and the south poles, respectively.
The original wave function near the poles are mapped to the Zeeman sublevels with $F_z=\mp F$, where the phase singularity are removed and the artificial vortices disappear.

\begin{widetext}

\begin{figure}[htpb]
  \centering
  \includegraphics[width=.950\linewidth]{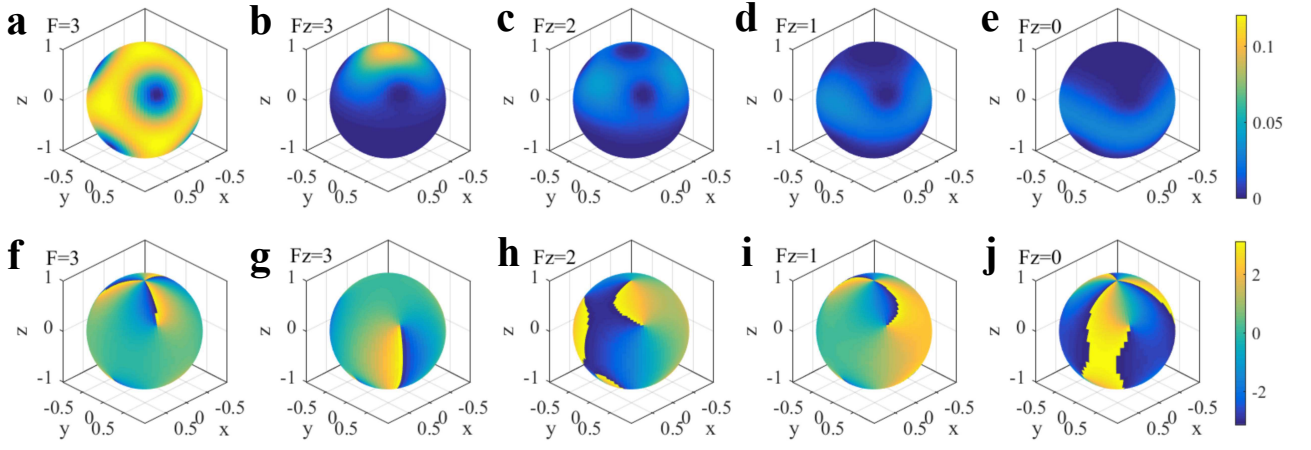}
  \caption{Density (upper panels) and phase (lower panels) portraits on sphere when $gR^2=50$ for $F=3$ and different Zeeman sublevels $F_z=-3$ ,$-2$ ,$-1$ ,and $0$. The two coreless vortices with $S=\mp F$ on the north and the south poles shown in (a) and (f) are due to the gauge effects we select, which can be removed by mapping the wave function to the usual Zeeman sublevels with different $F_z$, as shown in (b)-(e) and (g)-(j). For given $F_z$, there exists a gaint vortex with $\mp F+F_z$ units of circulation around the poles.  }
  \label{fig:Q=3all}
\end{figure}

\section{Dipole-dipole interaction on spherical surface}
In the lab, the long-range dipole-dipole interaction between dipolar atoms can be written as \cite{kawaguchi2012spinor}
\bea
U_{dip} (\vec{r}_1,\vec{r}_2) &=&    \frac{\vec{\mu}_1\cdot\vec{\mu}_2-3(\hat{r}_{12}\cdot\vec{\mu}_1)(\hat{r}_{12}
\cdot\vec{\mu}_2)}{r_{12}^3}  \nn \\
&=& \frac{1}{|\vec{r}_{12}|^3} \sqrt{\frac{6\pi}{5}} \sum_{m=-2}^{2} (-1)^m Y^{-m}_2(\hat{\Omega}_{12})\Sigma^m_2(\vec{\mu}_1,\vec{\mu}_2)
\eea
where ${r}_{12}=|\vec{r}_1-\vec{r}_2|$, $\hat{r}_{12}=(\vec{r}_1-\vec{r}_2)/r_{12}$, and $\hat{\Omega}_{12}$ is the orientation of $\vec{r}_{12}$. $Y_2^m(\hat{\Omega}_{12})$ is the usual spherical harmonics. $\Sigma^m_2$ is a rank-2 spherical tensor whose explicit form reads
\bea
\Sigma^{\pm 2}_2(\vec{\mu}_1,\vec{\mu}_2) &=& - \mu_{1,\pm}\mu_{2,\pm}, \nn \\
\Sigma^{\pm 1}_2(\vec{\mu}_1,\vec{\mu}_2) &=& \pm (\mu_{1,\pm}\mu_{2,z}+\mu_{1,z}\mu_{2,\pm}), \nn \\
\Sigma^0_2(\vec{\mu}_1,\vec{\mu}_2) &=& \frac{1}{\sqrt{6}} (\mu_{1,+}\mu_{2,-}+\mu_{1,-}\mu_{2,+} - 4 \mu_{1,z}\mu_{2,z} )
\eea
with $\mu_{\pm}=\mu_x \pm i \mu_y$.  In the rotating frame, the dipolar interaction becomes $U^{\dag} U_{dip} U$. Since $\vec{\mu} \propto \vec{F}$, only the spin part is modified. Using $U^{\dag} \mu_{\nu} U = \mu_{\nu} e^{i \nu \omega_L t} $ with $\nu=\{\pm,0\}$, we have
\bea
U^{\dag} U_{dip} U = \frac{1}{{r}_{12}^3} \sqrt{\frac{6\pi}{5}} \sum_{m=-2}^{2} (-1)^m Y^{-m}_2(\hat{\Omega}_{12})\Sigma^m_2(\vec{\mu}_1,\vec{\mu}_2) e^{im\omega_L t}.
\eea
For $m \neq 0$, the relevant interaction oscillates with frequency $m\omega_L$, which is much larger then typical energy scale defined by the trap $\omega_T$. The final effective interaction can then be written as
\bea
 U^{(e)}_{d} = \frac{1}{{r}_{12}^3} \sqrt{\frac{6\pi}{5}} Y^0_2(\hat{\Omega}_{12})\Sigma^0_2(\vec{\mu}_1,\vec{\mu}_2).
\eea
\end{widetext}
For hedgehog-type magnetic dipoles, this effective interaction is highly anisotropic around the spherical surface, and may result in instability of the condensates, as outlined in the main text.
When the condensates are trapped near the spherical shell, the approximated wave function can then be written as $\phi(\vec{r})=h(r)f(\hat{\Omega})$ with the density profile $n(\vec{r})=|\phi(\vec{r})|^2$. The dipolar interaction energy is estimated by
\bea
V_d=g_d\int d\vec{r} d \vec{r'} n(\vec{r}) \sum_{i,j} \Big[ \vec{m}_i(\vec{r}) U^{(e)}_{d} (\vec{r},\vec{r'})  \vec{m}_j(\vec{r'}) \Big ] n(\vec{r'}). \nn
\eea
Here $\vec{m}=(m_1,m_2,m_3)$ is a unit vector of the local dipole orientation, which is usually position-dependent in our case.


For hedgehog-like dipolar orientations considered in the paper, it is convenient to rewrite $\vec{m}=(m_1,m_2,m_3)^T=(\sin\theta\cos\phi,\sin\theta\sin\phi,\cos\theta)$ using the spherical basis defined by $(e_-,e_0,e_+)^T=\Lambda\cdot(m_1,m_2,m_3)^T$ with $\Lambda$ the unitary matrix
\bea
\Lambda=\left [
\begin{array}{ccc}
1/\sqrt{2} & -i/\sqrt{2} & 0 \\
0 & 0 & 1 \\
-1/\sqrt{2} & -i/\sqrt{2} & 0
\end{array}
\right ].
\eea
The dipolar interaction becomes divergent when ${r}_{12}\rightarrow 0$. To tackle it numerically, it is helpful to transform it into the momentum space. Since $(e_-,e_0,e_+)^T=\sqrt{\frac{4\pi}{3}}(Y^{-1}_1,Y^{0}_1,Y^{1}_1)^T$, and using the Fourier transformation of the dipolar interaction
\bea
U_{dip}(\vec{k}) &=& \mathcal{F}\left[\frac{Y^0_2(\hat{\Omega})}{r^3} \right] = -\frac{4\pi}{3} \big(\frac{1}{4}\sqrt{\frac{5}{\pi}} \big) \big[ \frac{3k_z^2}{|\vec{k}|^2} -1  \big]  \nn \\
&& = -\frac{4\pi}{3} Y^0_2(\hat{\Omega}_k),
\eea
we arrive
\bea
V_d=\left(\frac{4\pi}{3}\right)^2 g_d \int \frac{d\vec{k}}{(2\pi)^3} N^*_i(\vec{k}) T_{ij}(\vec{k}) N_j(\vec{k}),
\eea
where
\bea
&&N_j(\vec{k})=\int d \vec{r}\, n(\vec{r}) Y^j_l(\hat{\Omega}) e^{i\vec{k}\cdot\vec{r}},  \\
&&T_{ij}(\vec{k})= \alpha_{ij}[\Lambda^{\dag}U_{dip}(\vec{k})\Lambda]_{ij}=\alpha_{ij} Y^{i-j}_2(\hat{\Omega}_k),
\eea
and
\bea
\alpha = -\sqrt{\frac{4\pi}{5}}\left [
\begin{array}{ccc}
1 & 0 & 0 \\
0 & -2 & 0 \\
0 & 0 & 1
\end{array}
\right ].
\eea
The long-range characteristics of dipole-dipole interaction may result in erroneous results, which can be fixed by truncating the dipolar interaction within a radius $R_c$, as has been widely used in various literatures. Outside this regime, the dipolar interaction is negelected. This is reasonable due to the finite size of realistic physical setups.  This induces an additional truncated term in $T_{ij}(\vec{k})$ as $T_{ij}(\vec{k})=\alpha_{ij} T_c(k,R_c)Y^{i-j}_2(\hat{\Omega}_k)$ with
\bea
T_c(k,R_c)=1+\frac{3[kR_c\cos(kR_c)-\sin(kR_c)]}{(kR_c)^3}.
\eea

\begin{widetext}
Using the above formula, defining the radial and angular densities $n_R(r)=|h(r)|^2$ and $n(\hat{\Omega})=|f(\hat{\Omega})|^2=\sum_{lm}c_{lm}Y^m_l(\hat{\Omega})$, the nonlinear term in the GP equation related to the dipolar interaction can be written as
\bea
D(\hat{\Omega}) &=& \left(\frac{4\pi}{3}\right)^2\sum_{i,j=0,\pm 1} Y^{i*}_1(\hat{\Omega})\int dr r^2 n_R(r) \int \frac{d\vec{k}}{(2\pi)^3} T_{ij}(\vec{k}) N_j(\vec{k}) e^{-i\vec{k}\cdot\vec{r}} \nn \\
 &=& \frac{8}{3} \left(\frac{4\pi}{3}\right)\sum_{i,j=0,\pm 1} Y^{i*}_1(\hat{\Omega}) \sum_{l,m,l',m'}Y^m_l(\hat{\Omega}) \int d \hat{\Omega}_k \, Y^{m*}_l(\hat{\Omega}_k) T_{ij}(\hat{\Omega}_k) Y^{m'}_{l'}(\hat{\Omega}_k) E^{l'm'}_{1j} O_{ll'}
\eea
where we have also used the following expansions
\bea
e^{i\vec{k}\cdot\vec{r}}&=&4\pi \sum_{l,m} i^l j_l(kr) Y^{m*}_l(\hat{\Omega}) Y^{m}_l(\hat{\Omega}_k),   \\
N_j(\vec{k}) &=& \int d \vec{r} n(\vec{r}) Y^j_l(\hat{\Omega}) e^{i\vec{k}\cdot\vec{r}} = 4\pi \sum_{l'm'} i^{l'}\int dr r^2 n_R(r)j_{l'}(kr)Y^{m'}_{l'}(\hat{\Omega}_k) E^{l'm'}_{1j},   \\
O_{ll'}&=&i^{l'-l} \int dr dr' r^2 r'^2 n_R(r)n_R(r') j_l(kr) j_{l'}(kr')T_c(k,R_c),   \\
E^{l'm'}_{1j}&=&\sum_{l'm'} c_{l'm'}K^{l'm'}_{l''m'',1j}\mbox{ with } K^{l'm'}_{l''m'',1j}= \int d \hat{\Omega} \, Y^{m'*}_{l'}(\hat{\Omega}) Y^{m''}_{l''}(\hat{\Omega}) Y^{j}_{1}(\hat{\Omega}).
\eea
After simple algebra, we finally arrive
\bea
D(\hat{\Omega}) = \frac{8}{3} \left(\frac{4\pi}{3}\right) \sum_{i=0,\pm 1} \alpha_{ij} Y^{i*}_1(\hat{\Omega}) \sum_{l,m}Y^m_l(\hat{\Omega}) \sum_{l',m'} O_{ll'}K^{lm}_{l'm',20} E^{l'm'}_{1j}.
\eea
In this way, all the calculations related to the dipolar interactions can be solved using spherical harmonics. We also note that the relation between monopole harmonics and usual spherical harmonics can be obtained using the following identity \cite{wu1976dirac}
\bea
\mathcal{Y}^m_{l,F}\mathcal{Y}^{m'}_{l',F'} = \sum_{l''}(-1)^{l+l'+l''+F''+m''}\left [ \frac{(2l+1)(2l'+1)(2l''+1)}{4\pi} \right ]^{1/2}
\left( \begin{array}{ccc}
l & l' & l'' \\
m & m' & m''
\end{array} \right)
\left( \begin{array}{ccc}
l & l' & l'' \\
q & q' & q''
\end{array}\right) \mathcal{Y}^{-m''}_{l'',-F''}
\eea
with $m''=-m-m'$, $F''=-F-F'$, and $\left( \begin{array}{ccc}
l & l' & l'' \\ m & m' & m''
\end{array} \right)$ the $3$-j symbols.

\end{widetext}


\section{numerical details}

During our numerical calculation, we have set $\alpha=10$. This leads to the dimensionless radius of the spherical surface trap as $R\simeq \alpha =10$. The radial part of the wavefunction can be approximated as $h(r) \simeq  (\pi r^4)^{-1/4} \exp[-(r-R)^2/2]$, and is assumed to be fixed during the calculation. The GP equation is then solved using spectrum method based on spherical FFT \cite{healy2003ffts,fftaddress} with rescaled length $r' \sim r/R$. Using the rescaled length, the radial wavefunction is redefined as $\overline{h}(r') \simeq  (\pi \sigma^2 r'^4)^{-1/4} \exp[-(r'-1)^2/2\sigma^2]$ with $\sigma=1/R$. The corresponding interaction strengths are also modified as $g'=g$ and $g'_d=g_d\sigma^3$.

To obtain the density profiles within different Zeeman levels $|F,F_z\rangle$ for a given angular wave function $f(\hat{\Omega})$, we need to calculate the projection
\bea
\langle F,F_z|f(\hat{\Omega}) |\overline{F}\rangle_{\textbf{r}} &=& \langle F,F_z|f(\hat{\Omega}) S^{\dag}|F,-F\rangle_z \nn \\
&=& f(\hat{\Omega}) \langle F,F_z|e^{-iF_z\varphi} e^{-i F_y \theta}|F,-F\rangle_z \nn \\
&=& f(\hat{\Omega}) D^F_{F_z,-F}(\alpha=\theta,\beta=\varphi,\gamma=0),  \nn
\eea
which can then be solved using the standard Wigner $D$ functions.

\bibliographystyle{apsrev4-1}%
\bibliography{sphereV2}

\end{document}